\RequirePackage{fix-cm}
\documentclass[smallextended]{svjour3}       
\smartqed  
\usepackage{graphicx}
\begin{document}

\title{The resonance perturbations of the (39991) \textbf{Iochroma} family 
}


\author{Alexey Rosaev          
}


\institute{A. Rosaev \at
              Research and Educational Center ”Nonlinear Dynamics”, Yaroslavl State University, Yaroslavl, Russia \\
              \\
              \email{hegem@mail.ru}           
          }

\date{Received: date / Accepted: date}

\maketitle

\begin{abstract}

The dynamics of a very young compact asteroid cluster associated with asteroid 39991 Iochroma is studied. It is shown that Iochroma family lies between the two three body resonances 3J-1M-3  and 5J 3S-2. In this paper we have determined the position of these resonances and the boundary between them. 

{We have included the orbital elements approximations which may be useful in the 
future study of the dynamics of the family. Additionally, we report on a new 
candidate member, namely asteroid 2016 UT3.}

\end{abstract}
\


\section{Introduction}

Asteroids tend to group into so-called families or into associations of objects sharing similar orbits. Most of them are the results of very old (about 1 Gyrs) collisions between asteroids (\cite{SP2},\cite{NBC}).  Since the beginning of the twentieth century,  asteroid families and pairs have been the object of increasingly intensive studies.  

The detection of several asteroid families  and pairs with very recent formations  (about 1.5 Myr or less) (\cite{NVB},\cite{NV11}) in the past decades has generated a new and exciting development.  These discoveries are very important, because various collisional and dynamical processes have had little time to act on these families to alter their properties. 

The definition of close young asteroid pairs was given in paper \cite{NV2}. They propose the criterion of search for such pairs based on relative velocity estimations.   This criterion can also be applied in the case of compact young families. However, this is not a definition of asteroid pairs but rather an efficient method to identify candidate pairs. This limit should be used only as a first step towards identifying asteroid pairs or members of very young families.
Milani et al. \cite{Mil} proposed more robust identification criteria based on selecting candidates by successive filtering. However the problem requires further study.

Many cases of resonance perturbations of young families and pairs are known. A first example is the Datura family with its 9-16M resonance with Mars \cite{NV11}.  The chaotic orbits of the pair (49791) 1999~XF$_{31}$ and (436459) 2011~CL$_{97}$ may be explained by {the} 15-8M mean motion resonance with Mars \cite{P0}; the pair (7343) Ockeghem and (154634) 2003~XX$_{38}$ is in the 1J 1M-2 Jupiter-Mars-asteroid three-body resonance \cite{Du1}.

There are many publications devoted to studying the dynamics of resonance. 
We refer the reader to Morbidelli  \cite{Mo1}, Murray and Dermott \cite{MD}, 
Nesvorny and Morbidelli \cite{NM} and references therein.

Between the most recent papers, we note the paper by Eﬁmov and Sidorenko \cite{ES}, 
where interaction between two resonance modes is studied, and the paper by 
Petit \cite{Pt}, where an integrable model of three-body resonance is given. An 
important review concerning the main problems studied here is given by 
Carruba et al. \cite{Ca1}.

The main goal of this paper is to study the Iochroma cluster, an exciting example 
of a compact asteroid family, that orbits in the vicinity of two three-body 
resonances. Here we report our analysis of this group, including orbital elements 
approximations and resonances-related perturbations detections.

\section{Iochroma family main facts}

 The cluster associated with asteroid (39991) Iochroma (1998 HR37) was discovered 
 by Pravec and Vokrouhlicky \cite{PV}, who noted its belonging to the Nysa family. 
 Maybe, for this reason, the cluster is not included in the list of young asteroid 
 families \cite{NBC}. The cluster consists of 5 members with a relative velocity smaller 
 than 20 m/s, with the four secondaries being discovered in 2005-2008.
 
The pairs (39991) Iochroma and (349730) 2008 YU80, (39991) Iochroma and (340225) 2006 BR54 are listed as one closest, pair among the numbered asteroids with relative velocity $ < 0.5$  $m/s$ by Milani et al. \cite{Mi}. The close (and low velocity) encounter between (39991) Iochroma and (349730) 2008 YU80 at epoch about 7 kyr ago was studied by  Galad \cite{G}. He had shown that this epoch strongly depends on the perturbations by large asteroids.  

The information about the physical properties of the members of the cluster is very poor. From the photometric observations in Pravec et al. \cite{P0}, the color index in the Johnson-Cousins photometric system is consistent with an S type classification that is likely for this asteroid clusters. 

The values of the proper elements were taken from the AsDys site by Knezevic and Milani \cite{KM}. The values of the Lyapunov Characteristic Exponents are remarkably different between 1.87 $ Myr^{-1} $ and 11.3 $ Myr^{-1} $. This means that the orbits of the cluster have different stability for some reason, maybe due to resonance perturbations.

However, there is no detailed study of this group of minor planets. Even the age of the Iochroma cluster could not be estimated for a long time.  Only recently, Pravec et al. \cite{P0} give two estimations for the age of the family amounting to about 190  kyr and 140 kyr.

\begin{figure}
   \centering 
     \includegraphics[width=10.2cm]{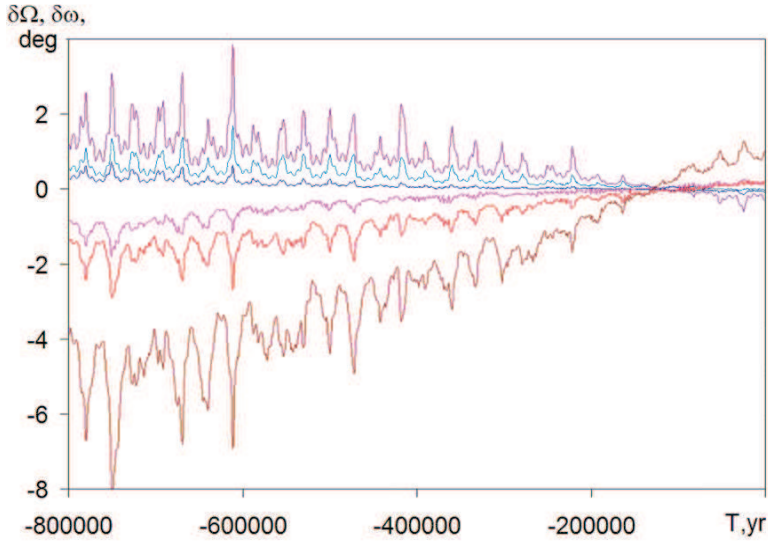}
     \caption{The convergence of the orbital elements  of Iochroma family; ((340225) 2006 BR54 is excluded). Here $\delta \Omega =\Omega_a-\Omega_{Iochroma},  \delta \varpi=\varpi_a  -\varpi_{Iochroma}$ }
\label{Conv}
 \end{figure}

\textbf{Age estimations.}
To confirm the values for the age, we made an independent age estimation based on our numerical integrations. The convergence of the angular orbital elements of the nominal orbits, when only large planets perturbations are considered, is very clear (Fig.\ref{Conv}). However, to obtain this result we need to exclude (340225) 2006 BR54 for our consideration.
For the formal estimation of the age of the cluster on the basis of nominal orbits analyses, we apply a method similar to our previous paper dealing with Hobson family \cite{RP2}. However, (39991) Iochroma is much larger than other members of the family and its orbit is evidently closer to the centre of the family.
 
 In summary, we obtain for the age of Iochroma family a value 170$\pm50$ kyr just between the two estimations \cite{P0}. Note, that we obtain our result without any non-gravity effects, based only on nominal orbits. 
 \  
\section{Numerical integration and approximation}

To study the dynamical evolution of the asteroids, the corresponding equations of the motion are numerically integrated over 800 kyr using the N-body integrator Mercury  \cite{Ch1} and the Everhart integration method \cite{E1}. 

We consider the gravitational effect of all planets + Ceres + Vesta + Juno + Pallas + Hygiea + Interamnia + Davida + Eunomia (CVJPHIDE) perturbations. The  mass values of the perturbing asteroids were taken from \cite{Ba}. The main model described above is aimed to maximally account all gravitational perturbations.

The resonance condition is:

\begin{equation}
 k_1\dot{\lambda}+k_2\dot{\lambda}_{Jupiter}+k_3\dot{\lambda}_{P}\approx 0  .
 \label{ra1}
 \end{equation}
 
Here $\lambda,\lambda_{Jupiter},\lambda_{P}$ are the longitudes of the asteroid, Jupiter and planet, respectively, and $k_1,k_2,k_3 $ are integers.
 
 In our procedure of resonance identification, we limit the set of possible combinations of the integers $k_1,k_2,k_3 $ by adopting the following conditions:
 
 \begin{equation}
  |k_1|, |k_2,|k_3| < K_{max} ;
  \label{ra2}
  \end{equation}
 we have used $K_{max}=15$, instead of $K_{max}=8$ as in paper \cite{SS}. 
  
For the nominal resonance position calculation, we use the values of the semimajor axis of the planets, averaged over the time of integration: 1.52368 AU for Mars, 5.20259  AU for Jupiter, 9.5549 AU for Saturn. To test our calculation of the resonant position $a_{pp} $, we compare it with the analytic $a_{A} $ and  numeric $a_{N} $ values in Nesvorny and Morbidelli \cite{NM}, Gallardo value ($a_{G} $) \cite{Ga} and Smirnov, Shevchenko paper ($a_{S} $) \cite{SS}.(see table \ref{tabResComp}).

We obtain a good agreement with the results of Nesvorny and Morbidelli \cite{NM}. The absolute values of the differences do not exceed 0.0005 AU. The Gallardo values are systematically less by about 0.001 AU. In general, this result confirms the validity of our method for the resonance computation.

However, we note a small difference with the numeric and analytic positions computed in \cite{NM}. Despite the small differences, the problem of the precision determination of the resonance position is still not solved.

\begin{table*}
\caption{{A comparison of the resonance position computations. }}
\begin{center}
\begin{tabular}{l l r r r r r} 
\hline 
 \multicolumn{2}{| c |}{Resonance (J:S:A) } & \multicolumn{1}{c |}{$a_{pp} $,AU}  & \multicolumn{1}{c |}{$a_{A} $,AU} & \multicolumn{1}{c |}{$a_{N} $,AU }&   \multicolumn{1}{c |}{$a_{G} $,AU } &   \multicolumn{1}{c |}{$a_{S} $,AU }  \\ 
  \hline \hline 
\multicolumn{1}{| c }{4-2-1} & \multicolumn{1}{l |}{}  & \multicolumn{1}{c |}{2.3976  }   &  \multicolumn{1}{c |}{ 2.3978}  &  \multicolumn{1}{c |}{2.3977}   & \multicolumn{1}{c |}{2.3967} & \multicolumn{1}{c |}{2.3985}    \\
\multicolumn{1}{| c }{4-3-1} & \multicolumn{1}{l |}{}  & \multicolumn{1}{c |}{2.6222}   &  \multicolumn{1}{c |}{2.6229}  &  \multicolumn{1}{c |}{2.6230}   & \multicolumn{1}{c |}{2.6211}  & \multicolumn{1}{c |}{2.6237}   \\
\multicolumn{1}{| c }{3-1-1} & \multicolumn{1}{l |}{}  & \multicolumn{1}{c |}{2.7527}   &  \multicolumn{1}{c |}{2.7527}  &  \multicolumn{1}{c |}{2.7525}   & \multicolumn{1}{c |}{2.7518} & \multicolumn{1}{c |}{2.7535}    \\
\multicolumn{1}{| c }{3-2-1} & \multicolumn{1}{l |}{}  & \multicolumn{1}{c |}{3.0790}   &  \multicolumn{1}{c |}{3.0794}  &  \multicolumn{1}{c |}{3.0790}   & \multicolumn{1}{c |}{3.0777} & \multicolumn{1}{c |}{3.0805}    \\
\hline
\end{tabular}
\end{center}
\label{tabResComp}
\end{table*}

To study the interaction of the considered family with the resonance and to determine the position of the resonance center (chaotic zone center), we computed the integration of the orbits of the asteroids with significant values of the Yarkovsky effect $(A_2=1*10^{-13}{AU/d^2})$ and different gravitation perturbations.

Due to the specific use of the integrator (Mercury), we needed to include the Yarkovsky non-gravitational force through $A_2$. In the Yarkovsky literature it is customary to report the strength of the Yarkovsky effect in terms of the semimajor axis drift rate $\dot{a}$.  In general, the value $\dot{a}$  is proportional to $A_2$, but can weakly vary with the eccentricity (see for example \cite{Fa1}). The used value of $A_2$ corresponds to $\dot{a}=10^{-3}$ AU/Myr, which is slightly larger than the value in \cite{Fa1}.

In our previous paper \cite{RP5}, we have shown analytically (and have proven numerically) that the frequencies of variation of $ \dot{\Omega}$  and $ \dot{\varpi}$ are related to the proper frequencies. For the perihelion longitude we have \cite{MD}:
\begin{equation}
\varpi = \varpi_0 + n \alpha \frac{m_p}{m_s} 2 \left[ 2C_1 t + \frac{e_p}{e \dot{\varpi}} C_3 \left( \sin \varpi - \sin \varpi_p \right)  \right] 
\label{omega2}
\end{equation}
or
\begin{equation}
\varpi = \varpi_0 + n \alpha \frac{m_p}{m_s} \left[ 2C_1 t + C_3 \frac{e_p}{e\dot{ \varpi}} \sin g_{1}t  \right]  ,
\label{omega3}
\end{equation}
where:
\begin{equation}
C_1=\frac{1}{8} \left[ 2\alpha \frac{d}{d \alpha} + \alpha^2 \frac{d^2}{d\alpha^2}  \right] b_{1/2}^{\left( 0 \right) } \left( \alpha \right) 
\label{C1}
\end{equation}

\begin{equation}
C_3= \frac{1}{4} \left[ 2 - 2\alpha \frac{d}{d \alpha} + \alpha^2 \frac{d^2}{d\alpha^2}  \right] b_{1/2}^{\left( 1 \right) } \left( \alpha \right)  .
\label{C3}
\end{equation}

Here $ b_{1/2}^{\left( k \right) } \left( \alpha \right)  $ are the Laplace coefficients and $\alpha = a/a_p$ is the ratio of the semi-major axis of the asteroid and the planet, $m_p$ is the mass of the perturbing planet, $m_s$ is the mass of the Sun. The index $p$ denotes the planet orbital elements.

In combination with the third order term, we have the following secular Lagrange equation for the node longitude:
\begin{equation}
\left[ \frac{d\Omega}{dt}\right]_{sec} = n\alpha \frac{m_p}{m_s} \left[ \frac{C_2}{2} + \frac{C_4 \sin i_p/2}{4\sin i/2} \cos\Omega\right] .
\label{dOmeha/dt-sec}
\end{equation}
Here:

\begin{equation}
C_2= - \frac{1}{2} \alpha b_{3/2}^{\left( 1 \right) } \left( \alpha \right) ,
\label{C2}
\end{equation}

\begin{equation}
C_4 = \alpha b_{3/2}^{(1)} + \frac{1}{2} \alpha \left( e^2 + e^2_p\right)\left[  1+2\alpha \frac{d}{d\alpha}+\frac{1}{2}\alpha^2 \frac{d^2}{d\alpha^2}\right] b_{3/2}^{(1)}  .
\label{C4new}
\end{equation}

A detailed derivation of this equation was given in our previous paper \cite{RP5}. The solution of this equation is:
\begin{equation}
\Omega_{sec} = n\alpha \frac{m_p}{m_c} \left[ \frac{C_2}{2}t + \frac{s_pC_4}{4\sin \frac{i}{2}} \sin\left( \dot\Omega t\right)  \right] .
\label{Omeha-sec}
\end{equation}

 \section{Results of the resonance perturbations search}
\begin{figure}
      \centering 
        \includegraphics[width=10.2cm]{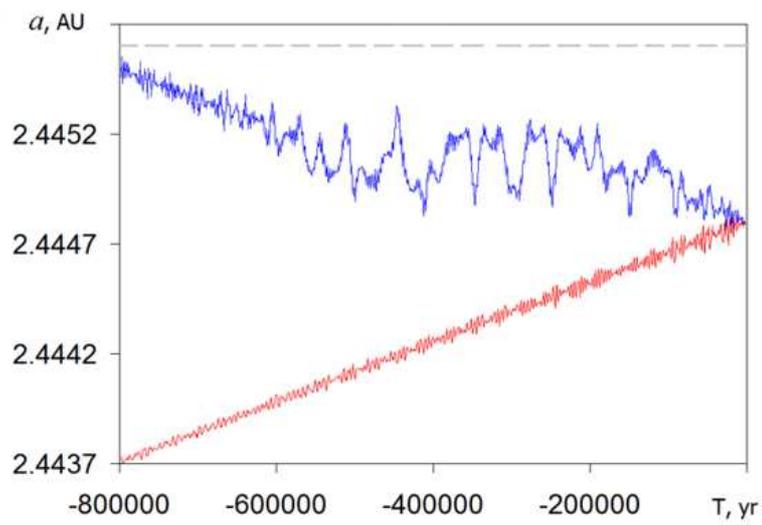}
        \caption{The semimajor axis evolution of 428243 (2006 YE19) with large Yarkovsky effect $A_2\pm 10^{-13} AU/d^2$: Mars and Jupiter perturbations. The calculated position of 3J-1M-3 resonance is marked by a dashed line.}
   \label{428243JM}
    \end{figure}

 Through our resonance search, we have obtained that the Iochroma family is unique: it lies between two close three-body mean motion resonances. The nearest 3-body resonance to the Iochroma family is: 3J-1M-3 at  2.445415 AU. When we use an integration with only Jupiter and Mars perturbations, and the Yarkovsky effect, we immediately detect this resonance. The chaotic zone center by numeric integration data is about 2.44502 AU for this resonance (Fig.\ref{428243JM}). The resonance argument can be written in the form:
 
 \begin{equation}
 \varphi=3\lambda-3\lambda_{Jupiter}-\lambda_{Mars} .
 \label{Lo1}
 \end{equation}
 
 Here $\lambda,\lambda_{Jupiter},\lambda_{Mars}$ are the longitudes of the asteroid, Jupiter and Mars, respectively.
 
\begin{figure}
      \centering 
        \includegraphics[width=10.2cm]{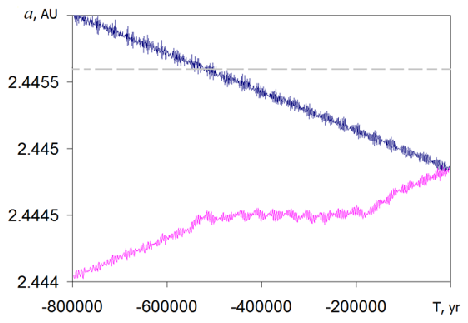}
        \caption{The semimajor axis evolution of 428243 (2006 YE19) with large Yarkovsky effect $A_2\pm 10^{-13} AU/d^2$: Saturn and Jupiter perturbations. The calculated position of 5J 3S-2 resonance is marked by a dashed line.}
   \label{428243JS}
    \end{figure}

 However, when we use only Jupiter and Saturn perturbations, and the Yarkovsky effect, we detect perturbations with center at about 2.4445 AU (Fig.\ref{428243JS}). This result corresponds to 5J 3S-2 resonance with a nominal position at 2.445661 AU.  The resonance argument in this case is:
  
  \begin{equation}
  \varphi=2\lambda-5\lambda_{Jupiter}-3\lambda_{Saturn} .
  \label{Lo2}
  \end{equation}
  
\begin{figure}
      \centering 
        \includegraphics[width=10.2cm]{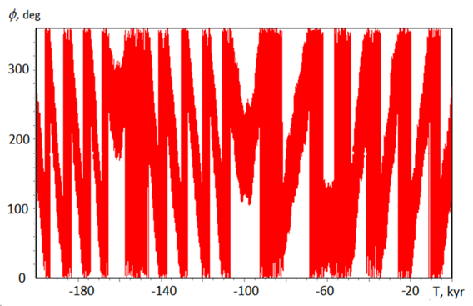}
        \caption{(513212) 2005 UU94 3J-1M-3 resonance argument evolution }
   \label{Arg}
    \end{figure}

  The asteroids of the Iochroma family are not in the exact resonance
 now. Therefore, it is likely that the resonance argument circulated almost all the time.
 In Fig \ref{Arg}, the behaviour of the 3J-1M-3 resonance argument for the
 asteroid (513212) 2005 UU94 (Eq.(\ref{Lo1})) is shown. We can see the short 
 periods of temporary resonance capture at 60, 100 and 160 kyr, which
 coincide with the epochs of crossing of the chaotic zone centre at 2.44505 AU 
 by this asteroid (see Section 6 below). A similar behaviour of the 
 3J-1M-3 resonance argument is obtained for (428243) 2006 YE19 and
 2016 UT3. At the same time, we have not found an epoch of critical
 argument libration for the Iochroma asteroids in 5J 3S-2 resonance. 
 
  Another close resonance has been found at  2.443322 AU (3J 8S-2), but it is too distant to provoke noticeable perturbations of the Iochroma family asteroids.
  
  However we note some shift of the 5J 3S-2 resonance relative to its nominal position. This raises a problem with the exact identification of this resonance and its interaction with the 3J-1M-3 resonance, as well as the  effect on the Iochroma family dynamics. We plan to consider this problem in detail in a future work.
 
  
  
  
    \  
    \
 
\section{Result of orbital elements approximations }
\subsection{The  precession related frequencies}

Here we present the results of our approximation of the orbital elements evolution of the members of the Iochroma cluster, based on the method described in \cite{RP5}. We pay attention to the eccentricity and inclination evolution. In addition to the classical Fourier series, we use the approximation by the expression (4) in \cite{RP5}:
\begin{equation}
E_i=E_{i00}+\sum_{k=1}^{N}  c_{ik} \cos \left( \omega_{ik}t+ \Phi_{ik} \right) \ .
\label{Ei_new_new}
\end{equation}
Here, the coefficients $c_{ik}$ are the amplitudes, $\omega_{ik}$ are the frequencies and $\Phi_{ik}$ are the phases.
In this paper, we take $N$=2. For eccentricity and inclinations in explicit form, we have:

\begin{equation}
e =e_{0}+a_{1e} \sin(w_{1e}t+\phi_{1e})+...
\label{e2}
\end{equation}
\begin{equation}
i =i_{0}+a_{1i} \sin(w_{1i}t+\phi_{1i})+...
\label{i2}
\end{equation}
  In the present paper, we assume the following approximate expressions for the angular elements:
 \
 \begin{equation}
 \Omega =\Omega _{0}+st+a_{s} \sin(s_{1}t+\phi_s)+...
 \label{Omega4}
 \end{equation}
 \begin{equation}
 \varpi =\varpi_{0}+gt+a_{g} \sin(g_{1}t+\phi_g)+...
 \label{omega4}
 \end{equation}
 \
 In conclusion, we obtain the equations (\ref{omega2}) and (\ref{Omeha-sec}). 
 
 To eliminate the short periodic perturbations, we have applied the averaging in a 1000-yr window before using the Fourier approximation and the expressions (\ref{Ei_new_new}) above. We have considered all large planets +CVJPHIDE perturbations.

The results of our approximations are shown in tables \ref{tabamplitnode} - \ref{tabamplitperih}. The asteroid name is in the first column, the synthetic proper frequencies $s $ and $ g$ from the AstDys website in the second columns, the corresponding mean frequency obtained using our numeric integrations (with index pp) in the third columns and the values of the frequency  ($\omega_{i1}$ or $\omega_{e1}$ ) in arcseconds per year in the fourth columns. In the last columns of tables \ref{tabamplitnode} - \ref{tabamplitperih}, we provided the values of the frequency ($\omega_{i2}$ or $\omega_{e2}$ ) in arcseconds per year.

The precision of the eccentricity frequencies ($\omega_{e1}$ or $\omega_{e2}$ ) is  $\pm0.002$ arsec/yr, the precision of the inclination frequencies ($\omega_{i1}$ or $\omega_{i2}$) is $\pm0.01$ arsec/yr.

\begin{table*}[h]
\caption{The {frequencies} for the approximation of inclinations and node longitudes of (39991)Iochroma family.}
\begin{center}
\begin{tabular}{l l r r r r } 
\hline 
 \multicolumn{2}{| c |}{Asteroid } & \multicolumn{1}{c |}{$s_{syn}["/yr]$  } & \multicolumn{1}{c |}{$s_{pp} ["/yr]$ } & \multicolumn{1}{c |}{$\omega_{i1} ["/yr]$}  & \multicolumn{1}{c |}{$\omega_{i2} ["/yr]$} \\ 
  \hline \hline 
\multicolumn{1}{| c }{(39991)} & \multicolumn{1}{l |}{Iochroma}  & \multicolumn{1}{c |}{ -46.3827 }   & 
\multicolumn{1}{c |}{-46.3751 }   &  \multicolumn{1}{c |}{-46.3787 }   &  \multicolumn{1}{c |}{20.11 }  \\
\hline
\multicolumn{1}{| c }{(340225)} & \multicolumn{1}{l |}{2006 BR54}  & \multicolumn{1}{c |}{ -46.3823 }   & 
\multicolumn{1}{c |}{-46.3753 }   &  \multicolumn{1}{c |}{-46.3890 }   &  \multicolumn{1}{c |}{20.13}   \\
\hline
\multicolumn{1}{| c }{(349730)} & \multicolumn{1}{l |}{2008 YV80}  & \multicolumn{1}{c |}{-46.3845 }   & 
\multicolumn{1}{c |}{-46.3824 }   &  \multicolumn{1}{c |}{-46.3787 }   &  \multicolumn{1}{c |}{20.11 }  \\
\hline
\multicolumn{1}{| c }{(428243)} & \multicolumn{1}{l |}{2006 YE19}  & \multicolumn{1}{c |}{ -46.3851}   & 
\multicolumn{1}{c |}{-46.3824 }   &  \multicolumn{1}{c |}{-46.3807}   &  \multicolumn{1}{c |}{20.11 }  \\
\hline
\multicolumn{1}{| c }{(513212)} & \multicolumn{1}{l |}{2005 UU94}  & \multicolumn{1}{c |}{ -46.3840 }   & 
\multicolumn{1}{c |}{-46.3788}   &  \multicolumn{1}{c |}{-46.3807 }   &  \multicolumn{1}{c |}{20.15 }  \\
\hline
\multicolumn{1}{| c }{} & \multicolumn{1}{l |}{2016 UT3}  & \multicolumn{1}{c |}{ - }   & 
\multicolumn{1}{c |}{-46.3788}   &  \multicolumn{1}{c |}{ -46.3869 }   &  \multicolumn{1}{c |}{20.10 }  \\
\hline
\end{tabular}
\end{center}
\label{tabamplitnode}
\end{table*}
 \begin{table*}[h]
 \caption{The frequencies for the approximation of eccentricity and perihelion longitudes of (39991)Iochroma family. }
 \begin{center}
 \begin{tabular}{l l r r r r } 
 \hline 
  \multicolumn{2}{| c |}{Asteroid } & \multicolumn{1}{c |}{$g_{syn}["/yr]$  } & \multicolumn{1}{c |}{$g_{pp} ["/yr]$ } & \multicolumn{1}{c |}{$\omega_{e1} ["/yr]$}  & \multicolumn{1}{c |}{$\omega_{e2} ["/yr]$} \\ 
   \hline \hline 
 \multicolumn{1}{| c }{(39991)} & \multicolumn{1}{l |}{Iochroma}  & \multicolumn{1}{c |}{ 41.9146 }   & 
 \multicolumn{1}{c |}{41.8775}   &  \multicolumn{1}{c |}{41.9813 }   &  \multicolumn{1}{c |}{13.752}  \\
 \hline
 \multicolumn{1}{| c }{(340225)} & \multicolumn{1}{l |}{2006 BR54}  & \multicolumn{1}{c |}{ 41.9146 }   & 
 \multicolumn{1}{c |}{41.8978 }   &  \multicolumn{1}{c |}{41.9813 }   &  \multicolumn{1}{c |}{13.745}   \\
 \hline
 \multicolumn{1}{| c }{(349730)} & \multicolumn{1}{l |}{2008 YV80}  & \multicolumn{1}{c |}{41.9174 }   & 
 \multicolumn{1}{c |}{41.9112 }   &  \multicolumn{1}{c |}{41.9875 }   &  \multicolumn{1}{c |}{13.805 }  \\
 \hline
 \multicolumn{1}{| c }{(428243)} & \multicolumn{1}{l |}{2006 YE19}  & \multicolumn{1}{c |}{ 41.9505}   & 
 \multicolumn{1}{c |}{41.9220 }   &  \multicolumn{1}{c |}{41.9813}   &  \multicolumn{1}{c |}{13.788 }  \\
 \hline
 \multicolumn{1}{| c }{(513212)} & \multicolumn{1}{l |}{2005 UU94}  & \multicolumn{1}{c |}{ 41.9462 }   & 
 \multicolumn{1}{c |}{41.8954}   &  \multicolumn{1}{c |}{41.9937}   &  \multicolumn{1}{c |}{13.768 }  \\
 \hline
 \multicolumn{1}{| c }{} & \multicolumn{1}{l |}{2016 UT3}  & \multicolumn{1}{c |}{ - }   & 
 \multicolumn{1}{c |}{41.8860}   &  \multicolumn{1}{c |}{41.9896  }   &  \multicolumn{1}{c |}{13.737}  \\
 \hline
 \end{tabular}
 \end{center}
 
 \label{tabamplitperih}
 \end{table*}

We notice a very good match of all values. The longitude of the perihelion frequency is $g= 41.91$ arcsec/yr. The first eccentricity frequency is $\omega_{e1}$ =41.98 arcsec/yr. We can conclude that a perturbation with the perihelion precession frequency (proper frequency $g$) is always present in the eccentricity evolution, as well as that the orbital plane precession frequency (proper frequency $s$) is always present in the inclination evolution. By taking into account the results of \cite{RP5}, we can state that this conclusion is true for almost all asteroids in the main belt. 

It is necessary to say a few more words about table \ref{tabamplitperih}. We note the difference between analytic and synthetic proper frequencies. In our computations of $g$, we have taken into account the classical value of the Jupiter frequency $g_j=4.25749$ arsec/yr (see AstDys site, section ``Help.Codes of secular resonances'' \cite{ast}) and we obtained a good agreement  with the synthetic value from AstDys site. However, it is necessary to use the correct value $g_j$ for a better accuracy. This correction can be obtained by our perturbation model (CVJPHIDE). 

\subsection{The resonance related frequencies}

In this section we discuss about the $\omega_{e2}$  frequency, which appeared after the subtraction of the $\omega_{e1}$ perturbation from the Iochroma asteroids eccentricities. The long period in eccentricity is about 94.24 kyr (table 2). This period in the eccentricity perturbations of the Iochroma family members can be explained by resonance perturbations. 

In the widely used first fundamental model of resonance \cite{MD} (the so-called pendulum model), two solutions are present: a libration solution for the exact resonance and a circulation solution for orbits outside of the resonance (see for example Wisdom \cite{W}). Here we deal with a circulation solution. The examples of such solution are discussed in the book by Murray and Dermott \cite{MD}.

The period of the resonance perturbation (P) increases as asteroid approaches the exact resonance \cite{Sb}. For the period of perturbation in the case of three body resonance we have:

\begin{equation}
P=\frac{2\pi}{\left | b n_{p1}+c n_{p2}-d\left (n_{r}+\delta  \right ) \right |} ,
\end{equation}

\begin{equation}
\delta_{n}=\left | n-n_{r} \right |=\left |\frac{2\pi }{P}  \right | .
\end{equation}

Here $ n_{p1} $ ,$ n_{p1}, n $ are the mean motions of the first and second planets, and the asteroid, respectively, while $ n_{r} $ is the  mean motion in the exact resonance, $b, c, d$ are integers. For the semimajor axis in the exact resonance $ a_{r} $  we have:

\begin{equation}
\delta_{a}=\left | a-a_{r} \right |\approx-\frac{2 a_{r}}{3 n_{r}}\delta_{n} =-\frac{2 a_{r}}{3 n_{r}}\left |\frac{2\pi }{P}  \right | ,
\end{equation}

\begin{equation}
a_{r}\approx\frac{a}{\left ( 1\pm 2/3\delta _{n}/n_{r} \right )}=a\left ( 1\pm\frac{4\pi }{3d\left ( bn_{p1}+ cn_{p2}\right )P} \right )^{-1} .
\label{ar}
\end{equation}

It is easy to see that the frequency of resonance perturbations decreases linearly with the distance from the resonance in semimajor axis:

\begin{equation}
f=\frac{1}{P}\approx \frac{3dn_{r}}{4\pi a_{r}}\left | a-a_{r} \right |=\frac{3\left ( bn_{p1}+ cn_{p2}\right )}{4\pi a_{r}}\left | a-a_{r} \right |
\end{equation}

\begin{figure}
    \centering 
      \includegraphics[width=10.2cm]{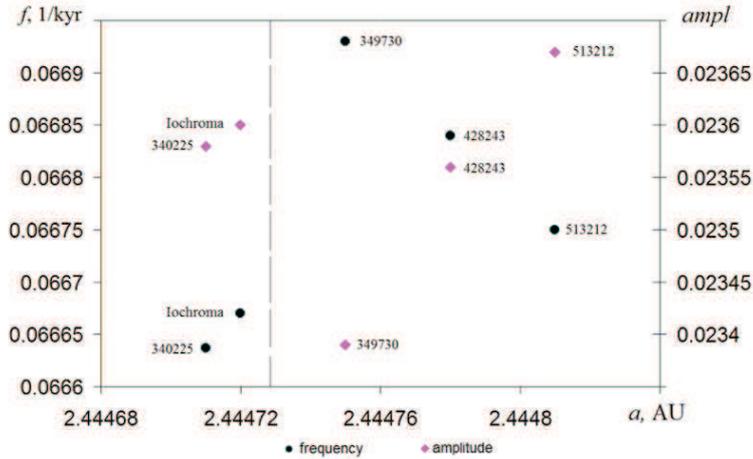}
      \caption{The amplitude (red) and frequency (black) of $\omega_{e2}$ versus the semimajor axis. A vertical dashed line marks (approximately) the boundary between the region in which the resonance is dominated (i.e. it has the largest amplitude of perturbations): left of it the 5J 3S-2 is dominated, right of it 3J-3M-1 is dominated.}
 \label{Res}
  \end{figure}

Three members of the Iochroma family: (349730) 2008 YV80, (428243) 2006 YE19,  (513212) 2005 UU94 have shown such dependence: their  frequency of resonance perturbations decreases linearly with the distance from the resonance (Fig. \ref{Res}). At the same time, the amplitude of the perturbations is linearly increasing during the resonance approach. 
  \
  
Other two asteroids, (39991) Iochroma and (340225) 2006 BR54, are close to the 5J 3S-2 Jupiter Saturn-asteroid three body resonance. 
Therefore, we can draw the corresponding resonances boundary at about $2.444735\pm0.000005 $ AU. 

As a result, we can draw the following important conclusions: 1) the second (long-periodic) perturbation in the eccentricity of the Iochroma family members is a resonance perturbation; 2) the sources of these perturbations are different for the different members of the cluster; 3) the position of the corresponding resonance coincided within good accuracy with other estimations.

\section{Search for new members of the Iochroma family}

 We have repeated the search for new orbits in the vicinity of the Iochroma cluster using the Lowell observatory catalogue \cite{Bo} on the date 15 June 2021; we found only one new candidate to be a member of the Iochroma family: 2016 UT3.
 To identify the new member, we follow two steps: 1) due to the compactness and small age of this family, we have provided a search in osculating orbital elements; 2) we have calculated the relative velocity using the method \cite{NV2}. The obtained values $v_{rel}<10$ m/s between 2016 UT3 and 39991 Iochroma are matched the criteria of a candidate for a close asteroid pair 
 
\begin{figure}
    \centering 
      \includegraphics[width=10.2cm]{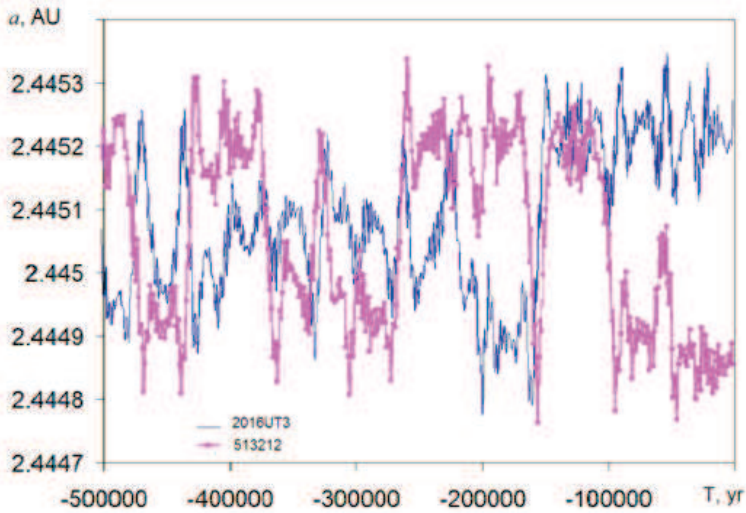}
      \caption{The semimajor axis evolution of (513212) 2005 UU94 (red) and 2016 UT3 (blue)}
 \label{NM}
  \end{figure}

 At present time, 2016 UT3 has moved on the opposite side of the 3J-1M-3 resonance and has a very chaotic orbit, like (513212) 2005 UU94. It is interesting that nominal orbits of both asteroids crossed the exact position of the resonance at the same epochs, - namely about 99 kyr and about 160 kyr ago, although in different directions (Fig.\ref{NM}). The second value is very close to the epoch of convergence of the orbital elements of the Iochroma cluster. 
This asteroid may be the key to understanding the origin of the Iochroma family, but only when its orbit will be improved.

  \

\section{Discussion and conclusions}

The dynamics of a very young compact asteroid cluster associated with asteroid 39991 Iochroma is studied. It is shown that the Iochroma family lies between two three body resonances 3J-1M-3 and 5J 3S-2.

As it is known, the interaction between close resonances can lead to their overlapping and as a consequence, to the appearance of chaos. Many theoretic studies of this problem are developed recently.  The Iochroma family is an interesting example of a system where the interaction between two resonances can produce a very complex dynamical behaviour. It is interesting to compare different theoretic models for studying this compact dynamical group of asteroids. 

In the present paper, we made the first step in this direction: we have estimated 
the position of both resonances and the boundary between them. We proposed a method 
to approximate the orbital elements and explained two periodic perturbations in eccentricity, 
one of which is related to resonances. Non-gravity perturbations complicate the 
problem; this effect must be considered in future studies.

Another problem requiring further study is the resonances' role in the origin of 
compact and young asteroid clusters and pairs.

\section{Compliance with Ethical Standards}
Conflict of Interest: The author declare that he has no conflict of interest.


\end{document}